# Analytical results in coherent quantum transport for periodic quantum dot

M. Mardaani[1], K. Esfarjani[2]

*Department of Physics, Sharif university of Technology, Tehran, 11365-9161, Iran*

**Email:** [1]mardaani@mehr.sharif.edu

[2]k1@sharif.edu

## Abstract

In this paper we have calculated electron transport coefficient in ballistic regime through a periodic dot sandwiched between uniform leads. We have calculated the Green's function (GF), density of states (Dos) and the coherent transmission coefficient (conductance) fully analytically. The quasi gap, bound states energies, the energy and wire-length dependence of the GF and conductance for this system are also derived.

## Introduction

The study of electron transport in Mesoscopic systems is one of the most fundamental problems in nanostructure physics. It has changed our understanding of transport phenomena in condensed matter systems [1,2,3]. In recent years there has been a growing interest theoretical study in electrical transport in quantum dots, molecular wires [4,6] and nanocrystals [5]. There has also been an increasing interest in the theoretical modeling of Molecular wire systems.

A Fundamental idea behind these advances was due to Landauer [3, 7] they stated that conductance is proportional to transmittance. We rely on Landauer theory as the basis for studying the conductance properties of quantum dot systems. This relates the lead-to-lead current to the transmission probability for an electron to scatter through the quantum dot.

## Model

In this section, we present our model and assumptions, and derive an analytic formula allowing one to calculate the conductance of a periodic dot analytically. The usual task of matrix inversion of the wire Hamiltonian is of the order of the number of orbitals (basis functions) in the wire. In this work GF and the transmission coefficient of a one-band chain can be calculated analytically.

*Periodic dot Hamiltonian attached to the two leads:*

The Hamiltonian of the periodic dot attached to the leads has the following form :

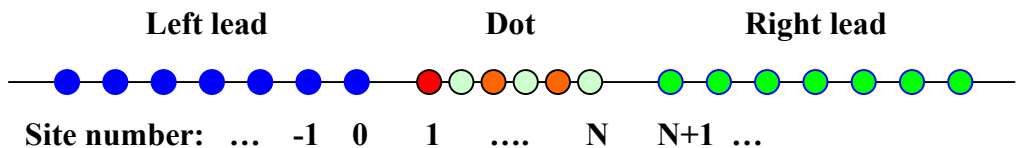

**Fig.1:** Periodic dot attached to the uniform leads.

$$H = H_D + H_L + H_R = \sum_z [\ E_z C_z^\dagger C_z - t_{z,z+1}(C_z^\dagger C_{z+1} + C_{z+1}^\dagger C_z)\ ] \qquad (1)$$

where

$$E_z = \begin{cases} E_D = E_0 + (-1)^{z+1} u & z = 1,\ldots, N \\ E_{0,L} = E_{0,R} = 0 & z < 1 \text{ or } z > N \end{cases} \ ; \ t_{z,z+1} = \begin{cases} t_L = t_R & z < 0 \text{ or } z > N \\ t_D & z = 1,\ldots, N-1 \\ t_{D,L} & z = 0, N \end{cases}$$

that in this equation, indexes $D,L,R$ refer to the dot, left lead and right lead respectively. The inverse GF of the dot differs from its isolated value only by the two self-energies added to the sites 1 and $N$.

$$[G^{-1}]_{ij} = [E + i\eta - E_i - \Sigma_L \delta_{i,1} - \Sigma_R \delta_{i,N}]\delta_{ij} - t_{D,i}(\delta_{i,j+1} + \delta_{i,j-1}) \quad \text{for } i,j = 1,\ldots,N \qquad (2)$$

where $\Sigma_L = \dfrac{t_{DL}^2}{t_L} e^{-i|k_L|a}$ ; $\Sigma_R = \dfrac{t_{DR}^2}{t_R} e^{-i|k_R|a}$ and $E = E_{L,R} - 2t_{L,R}\cos(k_{L,R}a)$

In this equation $a$ refers to the lattice constant.

In this case, we show that it can be inverted analytically. for the conductance only $G_{1N}$ is needed [8].

$$T(E) = tr(\Gamma_L G^{(+)} \Gamma_R G^{(-)}) = 4\,\text{Im}(\Sigma_L)\,\text{Im}(\Sigma_R)|G_{1N}|^2 \qquad (3)$$

After dividing by the hopping integral, the inverse of the GF of the periodic dot has the following form:

$$M(N, x-\alpha, y-\beta) = \begin{bmatrix} x-\alpha & -1 & 0 & \ldots & 0 & 0 & 0 \\ -1 & y & -1 & \ldots & 0 & 0 & 0 \\ 0 & -1 & x & \ldots & 0 & 0 & 0 \\ \ldots & \ldots & \ldots & \ldots & \ldots & \ldots & \ldots \\ 0 & 0 & 0 & \ldots & y & -1 & 0 \\ 0 & 0 & 0 & \ldots & -1 & x & -1 \\ 0 & 0 & 0 & \ldots & 0 & -1 & y-\beta \end{bmatrix}_{(N \times N)} \qquad (4)$$

$$D(N, x-\alpha, \beta) \equiv Det[M(N, x-\alpha, y-\beta)] \qquad (5)$$

where $x = \dfrac{z - E_D - u}{t_D}$ ; $y = \dfrac{z - E_D + u}{t_D}$ ; $\alpha = \dfrac{\Sigma_L(z)}{t_D}$ ; $\beta = \dfrac{\Sigma_R(z)}{t_D}$

The transmission coefficient $T(E)$ and $Dos(E)$ of a single chain can be written in the following form [see Appendix]:

$$G_{1N} = \dfrac{1}{D(N, x-\alpha, y-\beta)} ; G_{11} = G_{1N} D(N-1, y, y-\beta)$$

$$G_{NN} = G_{1N} D(N-1, x-\alpha, x) \qquad (6)$$

$$G_{2k,2k} = G_{1N} D(2k-1, x-\alpha, x) D(N-2k, x, y-\beta)$$

$$G_{2k+1,2k+1} = G_{1N} D(2k, x-\alpha, y) D(N-2k-1, y, y-\beta)$$

$$T(E) = 4\text{Im}(\Sigma_L)\text{Im}(\Sigma_R)|G_{1N}|^2 = \frac{4t_{DL}^2 t_{DR}^2 |\sin(ka)\sin(k'a)|}{t_D^2 t_L t_R |D(N, x-\alpha, y-\beta)|^2} \tag{7}$$

## *Results: Dependence on dot length and energy*

$G_{1N}$ has calculated following form:

$$\frac{1}{t_D G_{1N}} = (1 - \frac{\alpha}{x} - \frac{\beta}{y})D(N, x, y) + (\alpha\beta - \frac{\alpha}{x} - \frac{\beta}{y})D(N-2, x, y) \tag{8}$$

that

$$D(n, x, y) = \frac{(\xi + \sqrt{\xi^2 - 1})^{n+1} - (\xi - \sqrt{\xi^2 - 1})^{n+1}}{\sqrt{\xi^2 - 1}} \quad \text{and} \quad \xi = \frac{xy - 2}{2} = \frac{E^2 - u^2 - 2t_D^2}{2t_D^2}$$

For energy ranges outside the dot band (Quasi Gap) it can easily be seen that the behavior of $G_{1N}$ as a function of the energy is exponential since $|\xi| > 1$. A decay length may be obtained in this limit. In units of the lattice constant, it is the inverse of the coefficient of N in *log (G<sub>1N</sub>)*:

$$G_{1N} \longrightarrow \exp\left(-\frac{(N-1)a}{\lambda(E)}\right) \quad \text{if} \quad |\xi| > 1 \tag{9}$$

where $\xi = \frac{xy - 2}{2} = \frac{E^2 - u^2 - 2t_D^2}{2t_D^2}$ and $\lambda(E) = a / \log[|\xi| + \sqrt{\xi^2 - 1}]$

Likewise, the transmission coefficient T will have a similar behavior since it is quadratic in $G_{1N}$. Within the dot band, however, the behavior of $G_{1N}$ will be oscillatory with respect to N.

## *Analytical results for transport coefficient*

For periodic model we derived *T(E)* as following form in primitive range of energy (inside the dot energy band ) *T(E)* have oscillating behavior and in periodic dot with *N* atoms there are *N* peaks in the *T(E)*.
*T(E)* in the three energy region as following table:

**Table.1**: transmission coefficient for a periodic dot in the different energy range

| $\xi$ | $T(E)$ |
|---|---|
| $\xi < 1$ | $\dfrac{4\kappa^2 \sin^2(\theta)\cosh^2(\phi)}{\left((\cos\theta - \kappa\mu)\cosh(N+1)\phi - (\kappa^2\cos\theta - \kappa\mu)]\cosh(N-1)\phi\right)^2 + \sin^2(\theta)\left(\cosh(N+1)\phi + \kappa^2\cosh(N-1)\phi\right)^2}$ |
| $|\xi| \leq 1$ | $\dfrac{4\kappa^2 \sin^2(\theta)\sin^2(\phi)}{\left((\cos\theta - \kappa\mu)\sin(N+1)\phi + (\kappa^2\cos\theta - \kappa\mu)\sin(N-1)\phi\right)^2 + \sin^2(\theta)\left(\sin(N+1)\phi - \kappa^2\sin(N-1)\phi\right)^2}$ |
| $\xi > 1$ | $\dfrac{4\kappa^2 \sin^2(\theta)\sinh^2(\phi)}{\left((\cos\theta - \kappa\mu)\sinh(N+1)\phi + (\kappa^2\cos\theta - \kappa\mu)\sinh(N-1)\phi\right)^2 + \sin^2(\theta)\left(\sinh(N+1)\phi - \kappa^2\sinh(N-1)\phi\right)^2}$ |

where: $x = \dfrac{z-u}{t_D}$; $y = \dfrac{z+u}{t_D}$; $\xi = \dfrac{xy-2}{2}$; $\mu = \dfrac{x+y}{xy}$

and

$$\kappa = \dfrac{t_{DL}^2}{t_D t_L} \; ; \; \theta = \cos^{-1}(\dfrac{E}{2t_L}); \Phi = \begin{cases} \dfrac{1}{2}\cos^{-1}\xi & \text{if } |\xi| \leq 1 \\ \dfrac{1}{2}\ln\left|\xi + \sqrt{\xi^2 - 1}\right| & \text{if } |\xi| > 1 \end{cases}$$

and so $|E| \in \begin{cases} [\sqrt{u^2 + 4t_D^2}, \, 2|t_L|] & \text{if } \xi > 1 \\ [|u|, \sqrt{u^2 + 4t_D^2}] & \text{if } |\xi| \leq 1 \\ [0, |u|] & \text{if } \xi < -1 \end{cases}$

in the above expression we know: $E_{0L} = E_{0R} = 0$ and $u = \dfrac{1}{2}(E_{0D}^A - E_{0D}^B)$

For energy ranges outside the dot band (Quasi Gap) it can easily be seen that the behavior of T(E) as following form:

$$T(E,N) \to \left( \left| \dfrac{E^2 - u^2 - 2t_D^2}{2t_D^2} \right| + \sqrt{(\dfrac{E^2 - u^2 - 2t_D^2}{2t_D^2})^2 - 1} \right)^{-(N-1)} \qquad (10)$$

### *Bound states in the periodic attached to leads.*

There can be formation of bound states in the dot if there is an energy range within the bandwidth of the isolated dot, which falls below, or above all the energy levels of the leads. The bound state energies are the roots of the following equation (note that $\alpha$ and $\beta$ are functions of the energy $E$).

$$(1 - \dfrac{\alpha}{x} - \dfrac{\beta}{y})D(N, x, y) + (\alpha\beta - \dfrac{\alpha}{x} - \dfrac{\beta}{y})D(N-2, x, y) = 0 \qquad (11)$$

For simplicity if $\alpha$ and $\beta$ are equal we have:

$$F(E) = (1 - \alpha\mu)\sin(N+1)\Phi + \alpha(\alpha - \mu)\sin(N-1)\Phi = 0 \qquad (12)$$

where in eq.12

$$\alpha = \kappa \left( \left| \dfrac{E}{2t_{lead}} \right| - \sqrt{(\dfrac{E}{2t_{lead}})^2 - 1} \right)$$

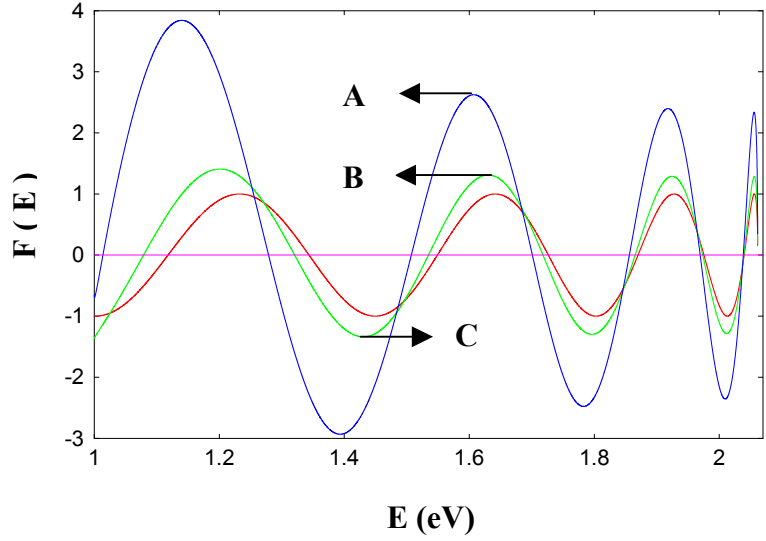

**Fig. 2**: Bound state energies of a 20-atom dot attached to two leads of periodic onsite energy. They are obtained by roots of the function F (E), The three curves are obtained for different dot-lead hoppings:

$$E_{0L} = E_{0R} = 0 \; ; |u| = \frac{1}{2}|E_{0A} - E_{0B}| = 0.5 \; ; t_D = 2t_L = 2t_R = -1$$

$$t_{DL} = t_{DR} = \{0:(A), -0.5:(B), -1.0:(C)\}$$

In Fig. 1, the function F has been displayed as a function the energy for a system where all onsite energies are equal to ±0.5 and the hopping of the dot is twice that of the leads. The three curves correspond to three different values of dot-to-lead hoppings. The bound state energies are the intersection of the curve of $F(E)$ with the $E$ axis. The curve of $F$ is even in energy and the same bound states also exist in the [-2,-1] $eV$ energy range.

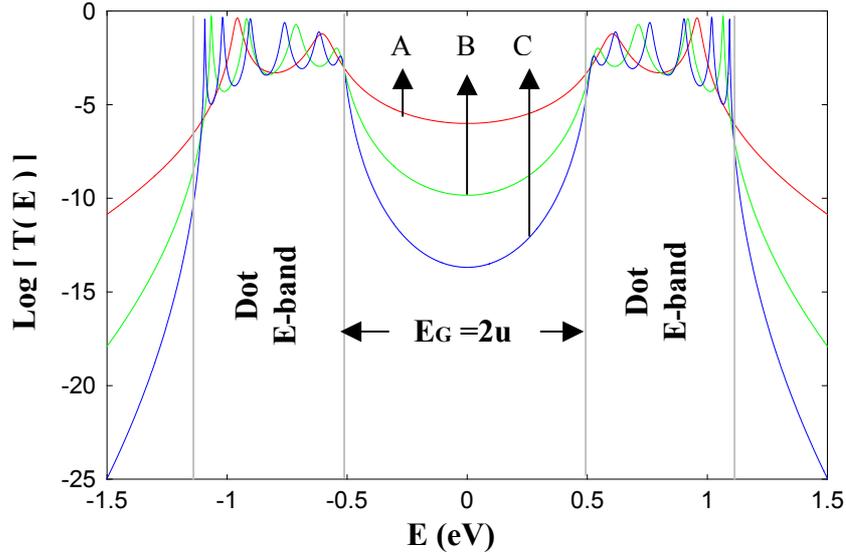

**Fig. 3:** this figure shows $\log[T(E)]$ vs. $E$ (eV)[Fermi energy] for the following specific parameter:
N=4(A),8(B),12(C); $t_L = t_R = 2t_D = -1; t_{DL} = t_{DR} = -0.25; u = 0.5$

In Fig. 2 there are three regions. In the band region of the N atom dot, there are N peaks that symmetrically distributed .In this region the transmission coefficient has oscillating behavior, but in the quasi gap, it decays.

For ($N>>1$), In the half band case in the pseudo gap region, the transmission coefficient is zero .If we apply a uniform gate voltage on the dot, this graph will shift and system can become either metallic or insulator depending on the position of the Fermi level with respect to the dot band.

## Appendix

To compute the transport coefficient and Dos of in the periodic model, we first need its determinant:

$$D(N=2n, x-\alpha, y-\beta) = (1-\frac{\alpha}{x}-\frac{\beta}{y})D(2n,x,y) + (\alpha\beta-\frac{\alpha}{x}-\frac{\beta}{y})D(2n-2,x,y) \quad (A.1)$$

$$D(N=2k+1, x-\alpha, x) = (1-\frac{\alpha}{x})D(2k+1,x,x) - \frac{\alpha}{x}D(2k-1,x,x)$$

$$D(N=2k+1, y, y-\beta) = (1-\frac{\beta}{y})D(2k+1,y,y) - \frac{\beta}{y}D(2k-1,y,y)$$

$$xD(2k+1, y, y) = yD(2k+1, x, x) = D(2k, x, y) + D(2k+2, x, y)$$

where we have defined the matrix $M_{2P}$ as follows:

$$M(2p,x,y) = \begin{bmatrix} x & -1 & 0 & ... & 0 & 0 & 0 \\ -1 & y & -1 & ... & 0 & 0 & 0 \\ 0 & -1 & x & ... & 0 & 0 & 0 \\ ... & ... & ... & ... & ... & ... & ... \\ 0 & 0 & 0 & ... & y & -1 & 0 \\ 0 & 0 & 0 & ... & -1 & x & -1 \\ 0 & 0 & 0 & ... & 0 & -1 & y \end{bmatrix}_{(2P\times 2P)} \quad (A.2)$$

$D_N = Det\, M_N$ ; then $D_N$ satisfies the following recursion relation:

$D(N=2p,x,y) = (xy-2)D(2p-2,x,y) - D(2p-4,x,y)$; with $D_0 = 1$ and $D_{-2} = -1$.

The solution results in:

$$D(2p,x,y) = \begin{cases} (-1)^p \frac{\cosh(2p+1)\Phi}{\cosh\Phi} & \text{if} \quad xy < 0 \\ \frac{\sin(2p+1)\Phi}{\sin\Phi} & \text{if} \quad 0 \le xy \le 4 \\ \frac{\sinh(2p+1)\Phi}{\sinh\Phi} & \text{if} \quad xy > 4 \end{cases} \quad (A.3)$$

where $\Phi = \begin{cases} \frac{1}{2}\cos^{-1}\xi & \text{if} \quad |\xi| \le 1 \\ \frac{1}{2}\ln|\xi+\sqrt{\xi^2-1}| & \text{if} \quad |\xi| > 1 \end{cases}$ and $\xi = \frac{xy-2}{2} = \frac{(z-E_{0,D})^2 - u^2 - 2t_D^2}{2t_D^2}$

$$D_{2p} = \begin{cases} (1-\frac{\alpha}{x}-\frac{\beta}{y})(-1)^p \frac{\cosh(2p+1)\Phi}{\cosh \Phi} + (\alpha\beta-\frac{\alpha}{x}-\frac{\beta}{y})(-1)^{p-1} \frac{\cosh(2p-1)\Phi}{\cosh \Phi} & \text{if } \xi < -1 \\ (1-\frac{\alpha}{x}-\frac{\beta}{y}) \frac{\sin(2p+1)\Phi}{\sin \Phi} + (\alpha\beta-\frac{\alpha}{x}-\frac{\beta}{y}) \frac{\sin(2p-1)\Phi}{\sin \Phi} & \text{if } |\xi| \leq 1 \\ (1-\frac{\alpha}{x}-\frac{\beta}{y}) \frac{\sinh(2p+1)\Phi}{\sinh \Phi} + (\alpha\beta-\frac{\alpha}{x}-\frac{\beta}{y}) \frac{\sinh(2p-1)\Phi}{\sinh \Phi} & \text{if } \xi > 1 \end{cases}$$